\documentclass[9pt]{article}  \usepackage{times}
\usepackage{graphicx}

\usepackage{color}

\usepackage[round,numbers,sort&compress]{natbib} 

\usepackage{amsmath}
\usepackage{amssymb}
\usepackage{stackrel}
\usepackage{chemarrow}
\usepackage{graphicx}
\usepackage{textgreek}

\usepackage{booktabs}
\usepackage{dcolumn}
\usepackage{rotating}
\usepackage{multirow}

\renewcommand\appendix{\par
  \setcounter{section}{0}
  \setcounter{subsection}{0}
  \setcounter{figure}{0}
  \setcounter{table}{0}
  \renewcommand\thesection{Appendix \Alph{section}}
  \renewcommand\thefigure{\Alph{section}\arabic{figure}}
  \renewcommand\thetable{\Alph{section}\arabic{table}}
}

\begin{document}

\twocolumn[{\LARGE \textbf{Linear nonequilibrium thermodynamics of reversible periodic processes and chemical oscillations\\*[0.2cm]}}
{\large Thomas Heimburg\\*[0.1cm]
{\small Niels Bohr Institute, University of Copenhagen, Blegdamsvej 17, 2100 Copenhagen \O, Denmark}\\*[-0.1cm]

{\normalsize \textbf{ABSTRACT}\hspace{0.5cm} Onsager's phenomenological equations successfully describe irreversible thermodynamic processes. They assume a symmetric coupling matrix between thermodynamic fluxes and forces. It is easily shown that the antisymmetric part of a coupling matrix does not contribute to dissipation. Therefore, entropy production is exclusively governed by the symmetric matrix even in the presence of antisymmetric terms. In this paper we focus on the antisymmetric contributions which  describe isentropic oscillations with well-defined equations of motion. The formalism contains variables that are equivalent to momenta and coefficients that are analogous to inertial mass. We apply this formalism to simple problems with known answers such as an oscillating piston containing an ideal gas, and oscillations in an LC-circuit. One can extend this formalism to other pairs of variables, including chemical systems with oscillations. In isentropic thermodynamic systems all extensive and intensive variables including temperature can display oscillations reminiscent of adiabatic waves. 
\\*[0.3cm] }}
\noindent\footnotesize {$^{\ast}$corresponding author, theimbu@nbi.ku.dk  \,;\,  http://membranes.nbi.ku.dk }\\
\vspace{0.3cm}
]



\normalsize
\section{Introduction}

Thermodynamics is usually applied to describe the state of ensembles in equilibrium as a function of the extensive and intensive variables independent of time. Linear nonequilibrium thermodynamics is an extension of equilibrium thermodynamics that is used to describe the coupling of equilibration processes of the extensive variables in environments with large viscosity. It is characterized by a formalism first introduced by Onsager \cite{Onsager1931a, Onsager1931b} based on a linear coupling of fluxes of extensive quantities and forces related to intensive quantities. These forces are given as derivatives of the entropy. Onsager's method introduces time in terms the rates of change of extensive quantities on their path towards equilibrium. Thermodynamics and linear nonequilibrium thermodynamics are related through the fluctuations of the extensive variables \cite{Einstein1910} and the concept of microscopic reversibility \cite{Onsager1931b}. Fluctuations are proportional to thermodynamic susceptibilities \cite{Greene1951, Callen1952}. E.g., the heat capacity is proportional to fluctuations in energy, the compressibility to fluctuations in volume \cite{Heimburg1998} and the capacitive susceptibility to fluctuations in charge \cite{Heimburg2012}. Fluctuation lifetimes are related to equilibration rates \cite{Onsager1931b}. Nonequilibrium thermodynamics is not typically applied to processes with inertia, which is rather the realm of mechanics. Among many other applications, mechanics is used to describe problems where point masses move in time within a potential without any friction, i.e., systems that are not obviously ensembles. Such systems are characterized by Hamilton's equations of motion. However, one can construct examples where both thermodynamics and mechanics lead to equivalent descriptions. 

\subsection*{Coupled gas containers}
\begin{figure}[ht]
	\centering
	\includegraphics[width=8cm]{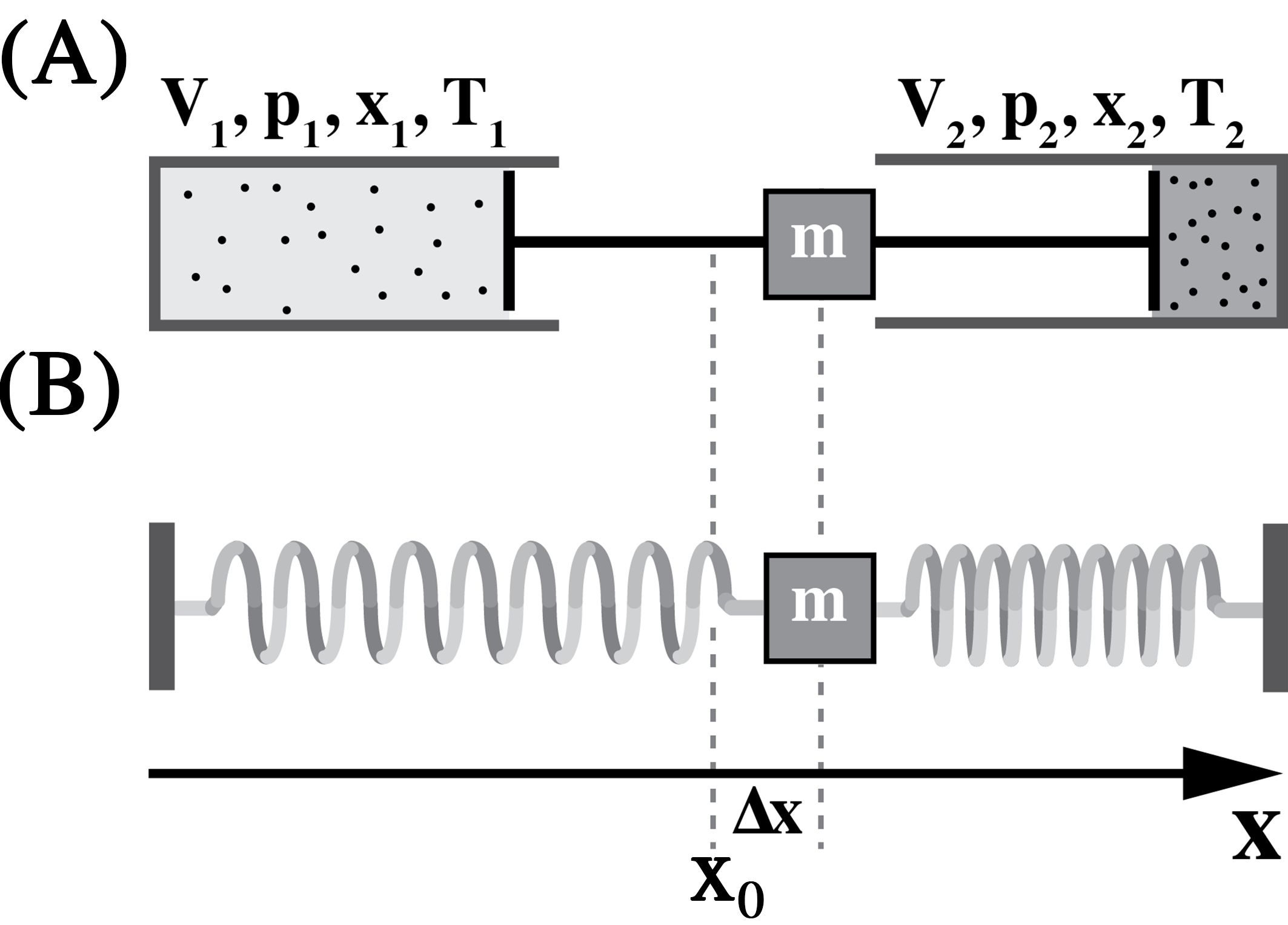}
	\caption{(A) Two reservoirs $1$ and $2$ containing an ideal gas, which are coupled by a piston with mass $m$. (B) Two springs with spring constant $K$ attached to a mass $m$. }
	\label{Figure_01}
\end{figure}

As an example, let us consider two identical containers, each filled with an adiabatically shielded ideal gas. They are coupled by a piston with mass $m$ (Fig. \ref{Figure_01}\,A). We further assume that the motion of the piston is frictionless and that no heat conduction occurs along the piston. The two gas volumes are thermodynamic ensembles, and their states are characterized by their volume, pressure and temperature. At the resting position of the system the entropy of the total gas is at maximum, and volume, pressure, and temperature of the two containers are given by $V_1=V_2=V_0$, $p_1=p_2= p_0$, and $T_1=T_2=T_0$. On can define a force $F=p\cdot A_0$ and a position $x=V/A_0$, where $A_0$ is the constant cross-section of the gas container.  If the position of the piston is brought out of equilibrium, the pressure of the two gases will be different and one obtains a force in the direction towards the equilibrium position. If the deviation from equilibrium, $\Delta x$, is small, the force in the piston is proportional to $\Delta x$. This is reminiscent of two coupled springs as shown in Fig. \ref{Figure_01}\,B.

The change in entropy of a mono-atomic gas due to changes in temperature and the position of the piston is given by
\begin{equation}
\label{Intro0_d1}
\Delta S =c_v\ln\left(\frac{T}{T_0} \right)+k\ln\left(\frac{x}{x_0} \right)
\end{equation}
where $c_v=(3/2) k$. The first term on the right hand side corresponds to a distribution of motional degrees of freedom, while the second part is the well-known entropy change of an isothermal compression that is solely of a configurational nature. During the adiabatic compression of an ideal gas the entropy stays constant, i.e. $\Delta S=0$. This leads to the fact that $T\cdot x^{2/3}$, $T^{5/3}\cdot F^{-2/3}$, and $F\cdot x^{5/3}$, are all time-independent constants. 

Since there are two forces, $F_1$ and $F_2$, from the two springs acting in opposite directions on the mass $m$, the total force is given by $\Delta F=F_1-F_2$. From the equations of state one obtains
\begin{equation}
\label{Intro0_a}
\Delta F=-\frac{10}{3}\frac{F_0}{x_0}\cdot \Delta x +O(\Delta x)^3\approx -K\cdot \Delta x\;,
\end{equation}
with a spring modulus $K=(10/3)\cdot F_0/x_0$ for the two springs combined, which is a constant with units [N/m]. In the ideal gas, $F_0\cdot x=NkT_0$ and
\begin{equation}
\label{Intro0_c}
K=\frac{10}{3 x_0^2}NkT_0 \;,
\end{equation}
where $N$ is the number of gas atoms in each container and $T_0$ is the temperature of the two containers in equilibrium (identical in both containers). This implies that the spring constant is proportional to temperature.

The force acting on the piston given by $\Delta F=-K\cdot \Delta x$ is equivalent to the notation in classical mechanics. The integral $V(\Delta x)\equiv -\int F dx=1/2\cdot K\cdot \Delta x^2$ is the mechanical potential which is proportional to $\Delta x^2$. It is in a minimum for $\Delta x=0$ (equilibrium) and equivalent to the maximum entropy. 

For $\Delta x \ne 0$, the temperatures in the two containers can be determined from $T\cdot x^{2/3}=$const. 
\begin{equation}
\label{Intro0_g}
\Delta T=T_2-T_1=-\frac{4\cdot T_0}{3}\frac{\Delta x}{x_0}+O(\Delta x)^3 
\end{equation}
and 
\begin{equation}
\label{Intro0_g2}
\Delta E=\Delta E_1+\Delta E_2=\frac{5}{3}Nk T_0\left(\frac{\Delta x}{x_0} \right)^2+O(\Delta x)^4 
\end{equation}
where temperature differences and positional differences are intrinsically coupled if both variables are free to change. 

If we do not fix the position of the piston, the above system will oscillate. The spring modulus can be determined from purely thermodynamic considerations given the assumption that each of the two gas containers are in equilibrium. This implies that the typical time scale of changes in position of the piston must be much smaller than the characteristic collision time of the gas particles at the given temperature and pressure. For instance, the mean velocity of Helium at room temperature is $\left\langle v\right\rangle =1245$m/s. At 1 bar pressure, the mean free path length is 12 \textmu m, the mean collision time is 9 ns, and the collision frequency is on the order of 100 MHz. For this reason, one can safely assume that in any practical realization of the above experiment each of the two gas volumes contains an equilibrated gas.
\begin{figure}[ht]
	\centering
	\includegraphics[width=8cm]{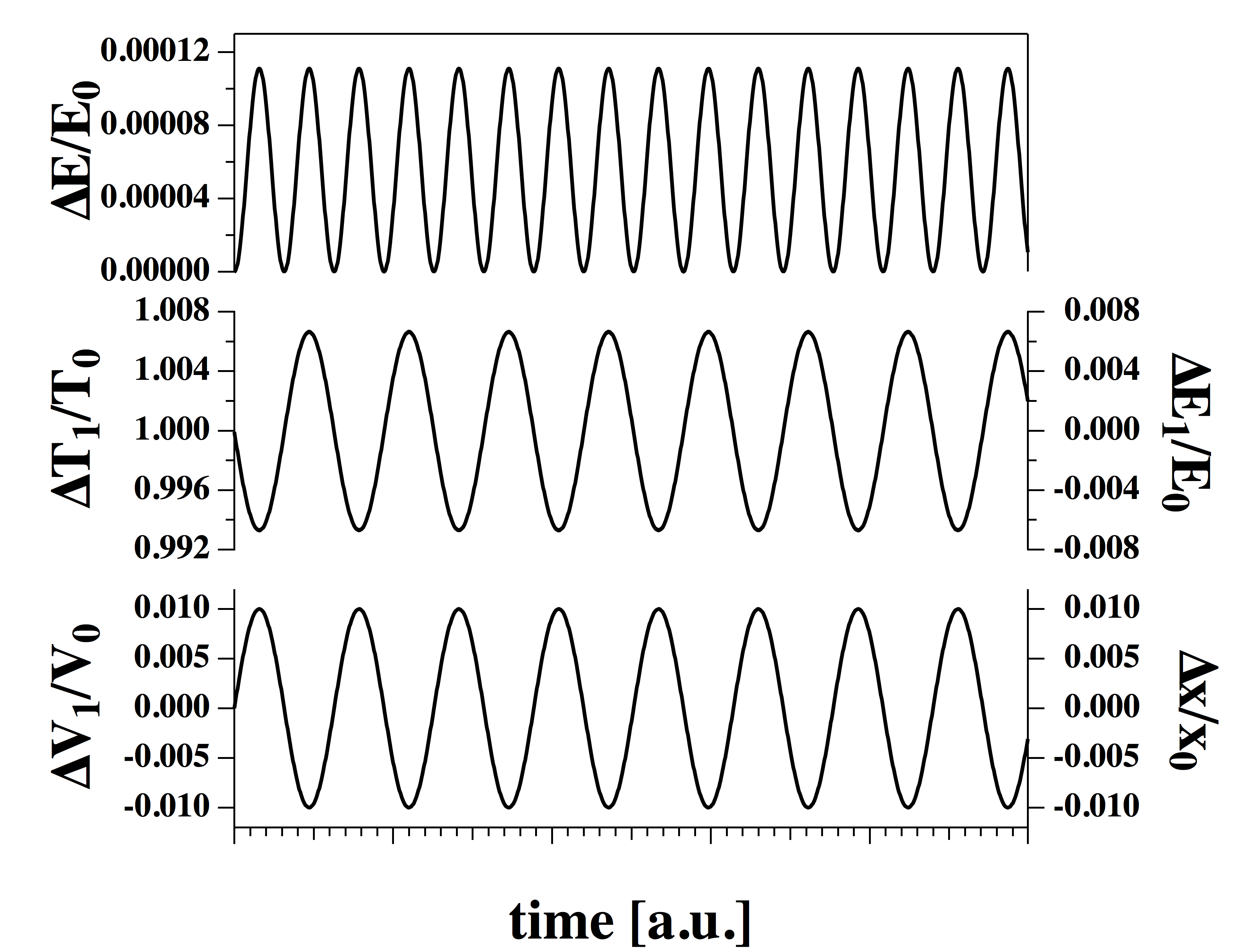}
	\caption{Oscillations of temperature, internal energy of the gas, volume and position in two coupled containers filled with an ideal gas. The frequency of the oscillation will depend on the mass of the piston, and the precise temperature and pressure of the ideal gas in equilibrium. The top panel shows the total internal energy of both containers. The amplitude of the oscillation is assumed being 1\% of the total gas volume.}
	\label{Figure_02}
\end{figure}
The frequency of oscillation is given by $\omega=\sqrt{(K/m)}$, where $m$ is the mass of the piston.  (The total mass of the gas is assumed to be much smaller than $m$ and has been neglected). The above considerations fail when the mass of the piston is comparable to the mass of the gas or the characteristic collision frequency of the gas is smaller than the frequency of the piston. Due to the equations of state, the gas containers will not only display oscillations in position but also display oscillations in temperature, pressure and internal energy of the gas (Fig. \ref{Figure_02}). Any adiabatic system with springs that are characterized by equations of state can be considered in complete analogy. 

In the presence of dissipation, macroscopic motion will disappear after some time and the internal energy of the gas will be distributed equally in the two containers. This could for example be realized when the piston conducts heat. This equilibration is not an isentropic process because the temperature increases in comparison to the zero-position of the oscillating piston by 
\begin{equation}
\label{Intro0_g3}
\Delta T_{max}=\frac{\Delta E_{max}}{3/2 \cdot N\,k}=\frac{10}{9}NkT_0\left( \frac{\Delta x_{max}}{x_0}\right) ^2+O(\Delta x)^4
\end{equation}
where $\Delta x_{max}$ is the maximum deviation of the piston from the equilibrium position. We see that dissipation leads to an increase in the overall temperature while in the absence of dissipation the temperature oscillates.

According to eq. (\ref{Intro0_d1}), the entropy increases by
\begin{equation}
\label{Intro0_g4}
\Delta S=c_v\cdot \ln\left( 1+\frac{\Delta T_{max}}{T_0}\right) =\frac{5}{3}Nk\left( \frac{\Delta x_{max}}{x_0}\right) ^2+O(\Delta x)^4
\end{equation}
The entropy is in its maximum when the amplitude of the oscillation is zero. It is interesting to note that except for a temperature-dependent prefactor, the change of entropy has a similar dependence on position as the mechanical potential $V(\Delta x)$. This seems natural considering that entropy carries the units [J/K] and the work displays units of [J]. We will therefore assume in the following that the entropy is the potential from which the forces can be derived - an assumption commonly made in non-equilibrium thermodynamics.
Thermodynamics describes the physics of the springs rather than that of the moving body and thus provides the origin of the potential energy.

A seemingly different class of oscillations are chemical oscillations such as the Belousov-Zhabotinsky (BZ) bromate reaction \cite{Zhabotinskii1964}, the iodine clock, \cite{Bray1921, Briggs1973} or yeast populations under stress conditions \cite{Teusink1996}. Such systems are also of a thermodynamic nature, and the variables that oscillate include the chemical potentials and the number of particles, respectively. The BZ-reaction is a chemical clock containing HBrO$_2$, Br$^-$, Ce$^{3+}$/Ce$^{4+}$, or O$_2$  as intermediates that oscillate in time \cite{Franck1978, Edelson1979, Vidal1980}. Such reactions are thought to originate from far-from-equilibrium processes \cite{Nicolis1977}. Typically, one describes them with a set of coupled non-linear rate equations containing auto-catalytic intermediate products as free variables. They are exemplified by well-known reaction schemes such as the Brusselator \cite{Nicolis1977, Kondepudi1998} or the Oregonator \cite{Field1974}. What such reaction schemes have in common is that temperature, pressure, the electrical potential and other thermodynamic variables not directly related to the concentrations, are not considered. \\
\begin{figure}[hbt]
	\centering
	\includegraphics[width=8cm]{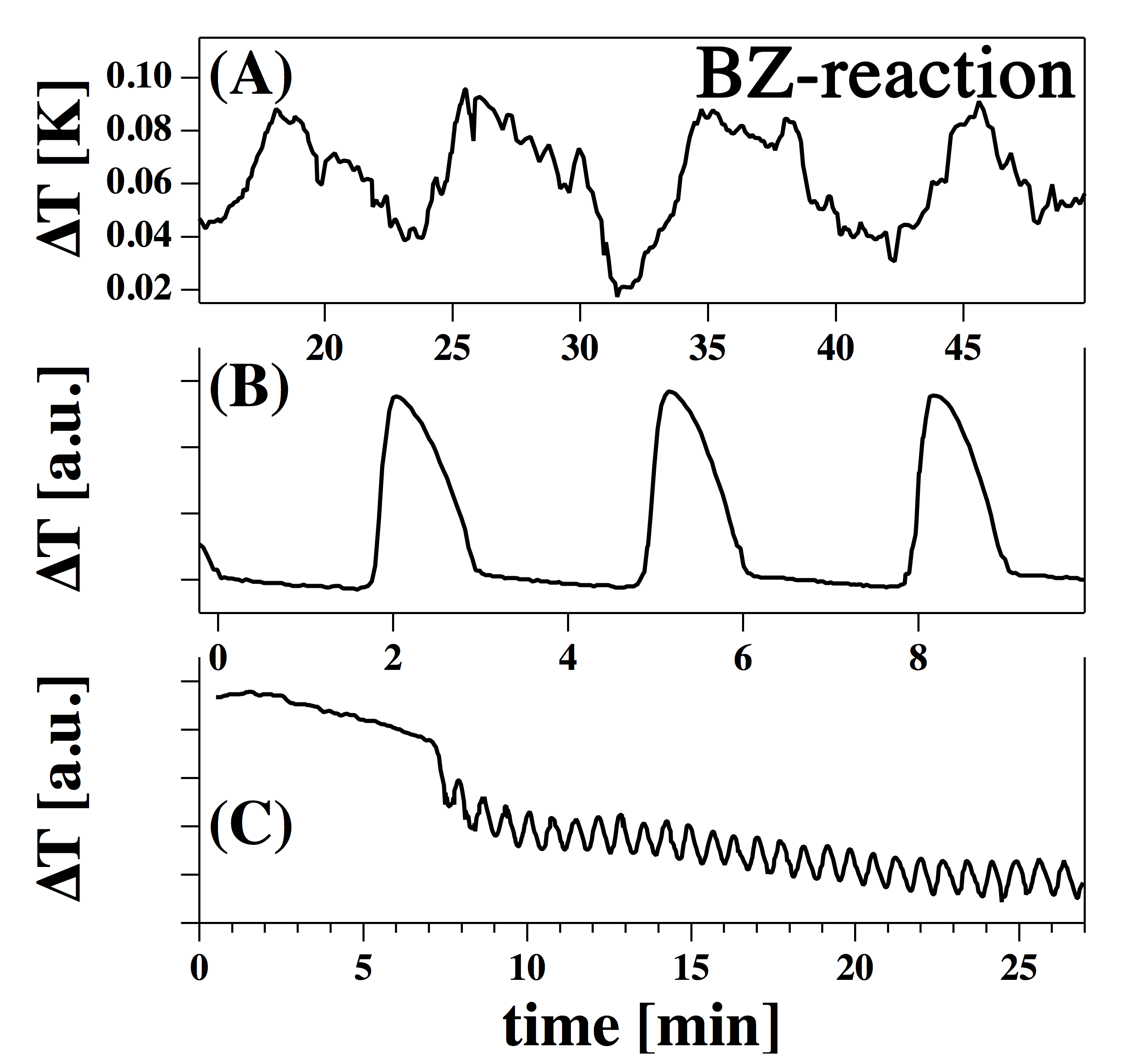}
	\caption{Temperature oscillations as a function of time in the Belousov-Zhabotinsky reaction and in nerves. (A) Adapted from \cite{Boeckmann1996} yielding up to 0.08 K in temperature variations. (B) Adapted from \cite{Franck1978} and (C) Adapted from \cite{Franck1971}. }
	\label{Figure_03}
\end{figure}
Interestingly, it is known that in some chemical systems such as the BZ-reaction the temperature oscillates in phase with the concentrations of the intermediates (Fig. \ref{Figure_03} (A-C), \cite{Franck1978, Franck1971, Koros1973, Boeckmann1996}). This is reminiscent of adiabatic oscillations such as the coupled gas containers in Fig. \ref{Figure_01}. Similar oscillations in  temperature or heat production rate have been reported in the Briggs-Rauscher reaction \cite{Lamprecht1987}. There are also biological systems with similar responses, e.g. yeast cells \cite{Teusink1996} or the action potential in nerves \cite{Abbott1958, Ritchie1985}, which that can produce periodic pulse trains and also display periodic temperature signatures. In the past, we have argued that the reversible changes in temperature found in nerves (Fig. \ref{Figure_03}\;D) indicate that the nerve pulse is an adiabatic pulse reminiscent of sound rather than a dissipative wave \cite{Heimburg2005c, GonzalezPerez2016}. The BZ-reaction shares similarities with the temperature response of nerves, i.e., it shares features of adiabatic processes.

In this paper, we describe a formalism reminiscent of Onsager's phenomenological equations \cite{Onsager1931a, Onsager1931b, Kondepudi1998}. In Onsager's formalism, only dissipative processes are considered, and the time scales are related to the fluctuation lifetimes. No inertia or momenta are taken into account. This is a good assumption for many processes with large friction, and Onsager chose to focus on such phenomena. The consequence is a symmetry in the coupling constants between fluxes and forces known as the reciprocal relations. However, the formalism cannot be applied to the motion of the two coupled gas pistons even though this is also a thermodynamic process that occurs in time. In this article we explore the possibility that mechanical oscillations and chemical oscillations are both related to adiabatic processes, and that they can be described with the methods of linear non-equilibrium thermodynamics that are modified such that they contain inertia. We show that in such oscillations, the system is both adiabatic and isentropic.

\section{Theory}
Einstein proposed to treat the entropy as a potential \cite{Einstein1910}. In harmonic approximation, the entropy can be expanded around that of the equilibrium state as
\begin{equation}
\label{Intro1}
S=S_0+\frac{1}{2}\sum_{ij}\underbrace{\left(\frac{\partial ^2  S}{\partial\xi_i \partial \xi_j}\right)_0}_{-g_{ij}} \xi_i \xi_j +... \approx S_0-\frac{1}{2}\sum_{ij}g_{ij}\xi_i\xi_j \;,
\end{equation}
where the $g_{ij}$ are the coefficients of a positive definite matrix with $g_{ij}=g_{ji}$ and $\det(\underline{\underline g})>0$ (all eigenvalues are positive). The variables are given by $\xi_i=(\alpha_i-\alpha_{0,i})$, where $\alpha_i$ is an extensive quantity (e.g.,  internal energy, volume, the number of particles of a particular species, charge, etc.), and $\alpha_{i,0}$ is the value of this variable in equilibrium. The consideration of the entropy as a potential led to the development of fluctuation-dissipation theorems pioneered by Greene and Callen \cite{Greene1951, Callen1952} and Kubo \cite{Kubo1966}.

Entropy production in a closed system evolving from a nonequilibrium state can be described by
\begin{equation}
\label{Intro2}
\frac{dS}{dt}\equiv\sigma=\sum_i\underbrace{\frac{\partial S}{\partial\xi_i}}_{\equiv X_i}\underbrace{\frac{\partial\xi_i}{\partial t_{}}}_{\equiv J_i} \equiv \sum_{ij}X_i J_i \;,
\end{equation}
where the $X_i=\partial S/\partial \xi_i$ are the thermodynamic forces and the $J_i=\partial \xi_i/\partial t$ are the conjugate thermodynamic fluxes of the variables $\xi_i$. 

Onsager's phenomenological equations \cite{Onsager1931a, Onsager1931b} expand the fluxes as linear combinations of the forces:
\begin{eqnarray}
\label{Intro3}
J_1&=&L_{11}^S X_1+L_{12}^S X_2+...\nonumber\\
J_2&=&L_{21}^S X_1+L_{22}^S X_2+...\\
&\vdots& \nonumber
\end{eqnarray}
or $\underline{J}=\underline{\underline{L}}^S\underline{X}$, where $\underline{\underline{L}}^S$ is a symmetric matrix. The symmetry is know as the `reciprocal relations'. Using eq. (\ref{Intro3}), the entropy production is given by
\begin{equation}
\label{Intro4}
\frac{dS}{dt}=\sum_{ij} L_{ij}^S X_i X_j =\sum_{ijkl}L_{ij}^S g_{ik}g_{jl}\xi_k\xi_l \;,
\end{equation}
where $X_i=-\sum_j g_{ij}\xi_j$ and $J_i=\sum L_{ij}^S X_j=$\linebreak
 $-\sum_{jk}L_{ij}^Sg_{jk}\xi_k$. 
\begin{figure*}[t!]
	\centering
	\includegraphics[width=16cm]{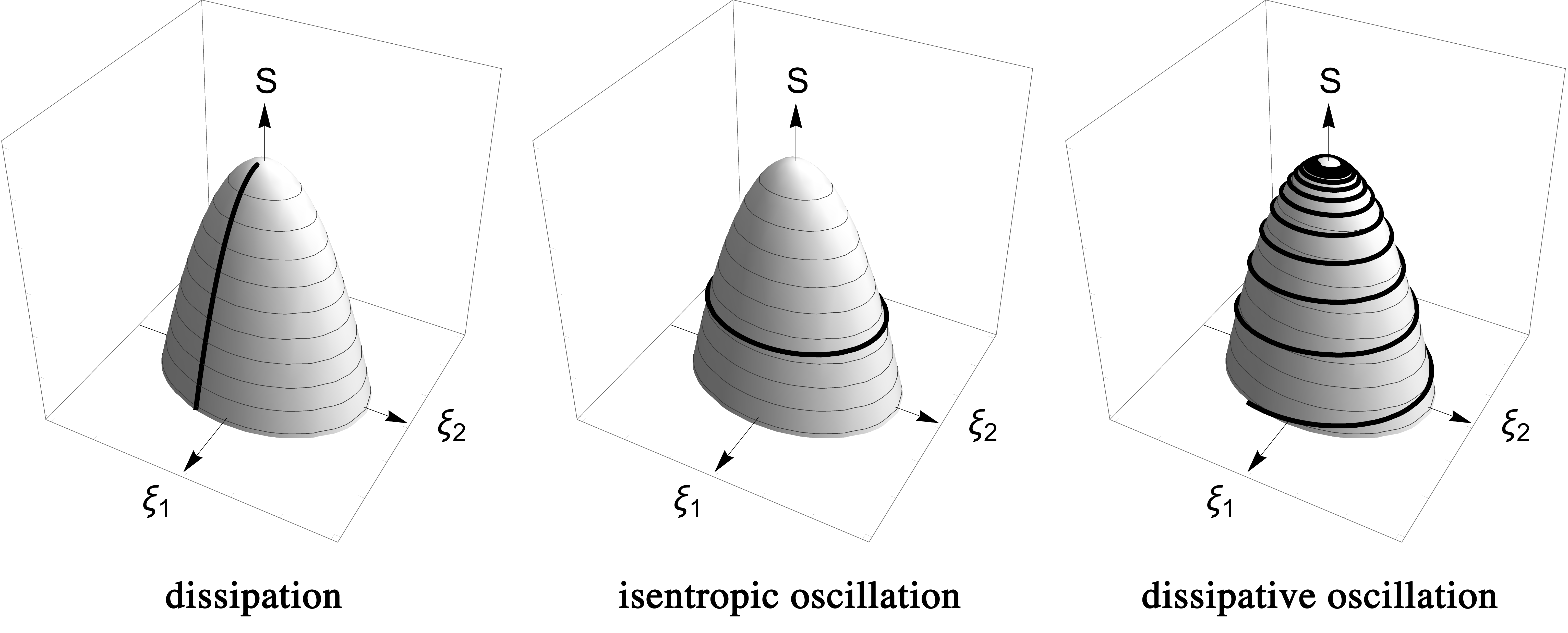}
	\caption{Schematic representation of dissipation (left), isentropic oscillations (center) and real processes with both oscillations and dissipation (right) in a harmonic entropy potential with two variables, $\xi_1$ and $\xi_2$.}
	\label{Figure_04}
\end{figure*}

\subsection*{Dissipation}
Onsager's decision not to consider the antisymmetric terms is based on the assumption that equilibrium fluctuations do not display a preferred direction in time, i.e., that detailed balance is obeyed \cite{Onsager1931b}. Onsager then postulated that a fluctuation and a macroscopic perturbation possess similar time evolutions and are both described by the phenomenological equation. This is plausible for over-damped systems where inertial forces are small compared to the forces created by thermal collisions. However, one can imagine nonequilibrium states of thermodynamic systems prepared such that they display a preferred direction in time. The example in Fig. \ref{Figure_01} is of this nature. In such systems, inertia is not generally small, and Onsager's argument may not apply. 

Any quadratic matrix $\underline{\underline{L}}$ can be written as a sum of a symmetric and an antisymmetric part, $\underline{\underline{L}}=\underline{\underline{L}}^S+\underline{\underline{L}}^A$, with
\begin{equation}
\label{T1.2}
L_{ij}^S=\frac{L_{ij}+L_{ji}}{2}\qquad\mbox{and}\qquad L_{ij}^A=\frac{L_{ij}-L_{ji}}{2} \;.
\end{equation}
The matrix $\underline{\underline{L}}^A$ describes isentropic processes because it can easily be seen that
\begin{equation}
\label{T1.3}
\sum_{ij}L_{ij}^A X_i X_j=0 \;.
\end{equation}
It is therefore a natural consequence which does not require any further justification that only the symmetric matrix $\underline{\underline{L}}^S$ contributes to dissipation (shown in Fig. \ref{Figure_04}, left), which was already discussed by Coleman and Truesdell \cite{Coleman1960} or Martyushev and Seleznev \cite{Martyushev2006}. Entropy production can be written more generally as $dS/dt=\sum_{ij}L_{ij}X_i X_j$ without making particular reference to the symmetry of the matrix $\underline{\underline{L}}$. As we will show, it is not generally justified to omit the antisymmetric terms.
The terms associated with the antisymmetric matrix describe processes that conserve entropy (i.e., oscillations, shown in Fig. \ref{Figure_04}, center) while the combination of the two leads to damped oscillations as schematically described by Fig. \ref{Figure_04} (right).
\subsection*{Oscillations}
In the following we explore the consequences of retaining only the antisymmetric terms. We will consider simple oscillations in a harmonic entropy potential. 

We assume systems with two variables, $\xi_1$ and $\xi_2$, with associated fluxes, $J_1$ and $J_2$, and two conjugated forces, $X_1$ and $X_2$. 
\begin{equation}
\underline{J}=\begin{pmatrix} 0 & L_{12}^A \\ -L_{12}^A & 0 \end{pmatrix} \underline{X}\;,
\end{equation}
There is only one coefficient, $L_{12}^A$, and the phenomenological equations read
\begin{eqnarray}
\label{T2.1}
\frac{d\xi_1}{dt}\equiv J_1 & = & L_{12}^A X_2 =-L_{12}^A(g_{21}\xi_1+g_{22}\xi_2)\nonumber\\
\frac{d\xi_2}{dt}\equiv J_2 & = & -L_{12}^A X_1=+L_{12}^A(g_{11}\xi_1+g_{12}\xi_2) \;.
\end{eqnarray}
$L_{12}^A$ may be positive or negative. The time derivatives of eq. (\ref{T2.1}) lead to
\begin{eqnarray}
\label{T2.2}
\frac{d^2\xi_1}{dt^2} & = &- L_{12}^A\left(g_{12}\frac{d\xi_1}{dt}+g_{22}\frac{d\xi_2}{dt}\right)\stackrel{ eq.(\ref{T2.1})}{=}
- \left(L_{12}^A\right)^2 \det(\underline{\underline g})\xi_1\nonumber\\
\frac{d^2\xi_2}{dt^2} & = & +L_{12}^A\left(g_{11}\frac{d\xi_1}{dt}+g_{12}\frac{d\xi_2}{dt} \right)\stackrel{ eq.(\ref{T2.1})}{=}
- \left(L_{12}^A\right)^2 \det(\underline{\underline g})\xi_2 \;,\nonumber\\
\end{eqnarray}
which displays periodic solutions with $\xi_{1}=\xi_{1,0} \cos(\omega t + \phi_1)$ and $\xi_{2}=\xi_{2,0} \cos(\omega t + \phi_2)$
with a frequency, $\omega$, given by $\omega^2=\left(L_{12}^A\right)^2 \det(\underline{\underline g})$. Thus, the antisymmetric part of the coupling matrix leads to oscillations.

The forces are linear functions of the fluxes of the extensive variables. Since the extensive variables oscillate, the conjugated thermodynamic forces also oscillate. This implies that in the isentropic case one also expects oscillations of the intensive variables such as temperature, pressure, electrical field and chemical potential.

\subsection*{Equations of motion for the isentropic case}
The entropy is given by
\begin{equation}
\label{T3.1}
S=S_0 -\frac{1}{2}g_{11}\xi_1^2-g_{12}\xi_1\xi_2-\frac{1}{2}g_{22}\xi_2^2=\mbox{const.}
\end{equation}
By using eq. (\ref{T2.1}) we obtain for the two forces $X_1$ and $X_2$
\begin{eqnarray}
\label{T3.2}
\frac{\partial S}{\partial \xi_1} =  -g_{11}\xi_1 - g_{12}\xi_2 & \stackrel{ eq.(\ref{T2.1})}{=} &-\frac{1}{L_{12}^A}\dot{\xi_2} \nonumber\\
\frac{\partial S}{\partial \xi_2} =  -g_{12}\xi_1 - g_{22}\xi_2 & \stackrel{ eq.(\ref{T2.1})}{=} & +\frac{1}{L_{12}^A}\dot{\xi_1} \;,
\end{eqnarray}
where $\dot{\xi_i}=d\xi_i/dt$. We will call eqs. (\ref{T3.2}) the thermodynamic equations of motion. 

For an arbitrary even number of variables, we find that
\begin{equation}
\label{T3.4}
\underline{\ddot\xi}=-\underline{\underline{L}}^A\cdot\underline{\underline{g}}\cdot\underline{\dot \xi}=\left(\underline{\underline{L}}^A\cdot\underline{\underline{g}}\right)^2\cdot\underline{\xi}
\end{equation}
with oscillatory solutions. The thermodynamic equations of motion are given by 
\begin{equation}
\label{T3.5}
\underline{X}=\underline{\frac{\partial S}{\partial \xi}} =\left(\underline{\underline{L}}^A\right)^{-1}\underline{\dot \xi} \;.
\end{equation}

\textbf{Simplification:} For simplicity we assume in the following that $\underline{\underline{g}}$ is a diagonal matrix meaning that the principal axes of the entropy potential align with the variables under consideration. In the case of two forces and fluxes, eq. (\ref{T2.1}) yields
\begin{equation}
\label{Examples1.2}
\xi_2=-\frac{1}{L_{12}^A g_{22}}\frac{d \xi_1}{dt} \;.
\end{equation}
Therefore, the second variable is proportional to the temporal variation of the first variable, i.e., it is reminiscent of a momentum. Eq. (\ref{T2.2}) becomes
\begin{equation}
\label{Examples1.3}
\frac{d^2\xi_1}{dt^2}  = - \left(L_{12}^A\right)^2 (g_{11}g_{22})\xi_1
\end{equation}
with oscillations of frequency $\omega^2=\left(L_{12}^A\right)^2 (g_{11}g_{22})$. The thermodynamic equations of motion (eq. (\ref{T3.2})) are given by
\begin{eqnarray}
\label{Examples1.6}
\frac{\partial S}{\partial \xi_1} =  -g_{11}\xi_1 &=&-\frac{1}{L_{12}^A}\dot{\xi}_2\nonumber\\
\frac{\partial S}{\partial \xi_2} = - g_{22}\xi_2 & = & +\frac{1}{L_{12}^A}\dot{\xi}_1 \;.
\end{eqnarray}

Eqs. (\ref{T3.2}), (\ref{T3.4}) and (\ref{Examples1.6}) display a formal similarity to Hamilton's equations of motion, which for one spatial variable $x$ with an associated momentum $p$, are given by
\begin{equation}
\label{T3.3}
\frac{\partial \mathcal{H}}{\partial x}   =  -\dot{p} \qquad \mbox{and} \qquad \frac{\partial \mathcal{H}}{\partial p}   =  +\dot{x} \;,
\end{equation}
where $\mathcal{H}$ is the Hamiltonian.

If dissipation is included, the phenomenological equations are given by
\begin{equation}
\underline{J}=\begin{pmatrix} L_{11}^S & L_{12}^A \\ -L_{12}^A & L_{22}^S \end{pmatrix} \underline{X}\;,
\end{equation}
where $L_{11}^S\cdot g_{11}=L_{22}^S\cdot g_{22}$. This situation is shown in Fig. \ref{Figure_04}, right.
\section*{Examples}
The above can be used to describe some very simple examples where the result is known.
\subsection*{Oscillating piston (or spring)}
Let us consider two coupled pistons with an associated mass, m, as given in Fig.\ref{Figure_01} (A). We assume that the position of the mass in equilibrium is given by $x_0$ and the deviation from equilibrium is $\Delta x$. Further, the pistons possess an adiabatic compression modulus $K$ that can be calculated from the adiabatic equations of state of an ideal gas as done in the introduction. (Two metal springs are conceptually equivalent but would possess different equations of state.)

Let us consider two variables, $\xi_1\equiv\Delta x$ and $\xi_2$, and that $g_{ij}$ is a diagonal matrix.

From eq. (\ref{Examples1.2}) it follows that
\begin{equation}
\label{Ex1.13}
\xi_2=-\frac{1}{L_{12}^A g_{22}}\frac{d \Delta x}{dt} \;.
\end{equation}

The entropy potential is therefore given by
\begin{equation}
\label{Ex1.19}
\Delta S=-\frac{1}{2}g_{11}\Delta x^2-\frac{1}{2g_{22} (L_{12}^A)^2}\left(\frac{d\Delta x}{dt}\right)^2
\end{equation}

In an equilibrium situation, the $\Delta x$ and $\Delta \dot{x}$ fluctuate independently of each other, and $\Delta S$ is not constant. For a well-defined pair $\Delta x$ and $\Delta \dot{x}$ the probability is given by
\begin{equation}
\label{Ex1.19b}
P=P_0\exp\left( \frac{\Delta S (\Delta x, \Delta \dot{x})}{k}\right) 
\end{equation}
If it is integrated over $\Delta \dot{x}$, it yields a Gaussian distribution for the fluctuations in position of the piston, 
\begin{equation}
\label{Ex1.19c}
P_x=\sqrt{\frac{g_{11}}{2\pi k} } \cdot \exp\left( -\frac{g_{11} \Delta x^2}{2k}\right) \equiv \sqrt{\frac{K}{2\pi kT}}\cdot \exp\left(-\frac{K x^2}{2 kT}\right)
\end{equation}
which is just the positional equilibrium fluctuations of an elastic spring. Here, $g_{11}\equiv K/T$ corresponds to the compression modulus (spring constant $K$) of the setup in Fig. \ref{Figure_01} (A). It is well-known from the fluctuation theorems that the compression modulus is related to the equilibrium fluctuations in position. 

When $P$ is integrated over $\Delta x$ instead, it yields a one-dimensional Maxwell distribution for the fluctuations in velocity of the piston
\begin{eqnarray}
\label{Ex1.19d}
P_{\dot{x}}&=&\sqrt{\frac{1}{2(L_{12}^A)^2\pi g_{22}k} }\cdot \exp\left( -\frac{\Delta \dot{x}^2}{2 g_{22}(L_{12}^A)^2 k}\right) \nonumber\\
&\equiv& \sqrt{\frac{m}{2\pi kT} }\cdot \exp\left(-\frac{m \dot{x}^2}{2 kT}\right)
\end{eqnarray}
which represents the equilibrium velocity fluctuations of the spring. Here, $1/g_{22}(L_{12}^A)^2\equiv  m/T$ corresponds to the mass attached to the piston in Fig. \ref{Figure_01} (A).
It is important to point out that the entropy $\Delta S$ responsible for the equilibrium fluctuations is the same as the one in which oscillatory motion takes place and that the elementary constants necessary to describe the oscillatory motion are already contained in the equilibrium fluctuations.

Entropy conservation leads to
\begin{equation}
\label{Ex1.19a}
\Delta S=-\frac{1}{2}g_{11}\Delta x^2-\frac{1}{2g_{22} (L_{12}^A)^2}\left(\frac{d\Delta x}{dt}\right)^2=\mbox{const.}
\end{equation}

Using the abbreviations introduced above, eq. (\ref{Ex1.19a}) can be rewritten as
\begin{equation}
\label{Ex1.19b2}
-T\Delta S=\frac{1}{2}K\Delta x^2+m \left (\frac{d\Delta x}{dt}\right)^2=\mbox{const.}\;,
\end{equation}
which corresponds to the equation for energy conservation in mechanics. In our examples, the energy in mechanics is equivalent to $-T\Delta S$ in thermodynamics, i.e., a free energy.
Eq. (\ref{Examples1.3}) yields
\begin{equation}
\label{Ex1.14}
\frac{d^2 (\Delta x)}{dt^2}  =  -\left(L_{12}^A\right)^2\left(g_{11}g_{22}\right)\Delta x \;.
\end{equation}
Thus, the oscillatory frequency is given by
\begin{equation}
\label{Ex1.21}
\omega^2=g_{11}g_{22}\left(L_{12}^A\right)^2\equiv\frac{K}{m} \;.
\end{equation}
The thermodynamic forces are given by
\begin{eqnarray}
\label{Ex1.23}
X_1 & = &  -g_{11}\Delta x\equiv -\frac{K\Delta x}{T}\qquad\mbox{and} \nonumber \\
X_2 & = &  -g_{22}\xi_2=+\frac{1}{L_{12}^A}\frac{d\Delta x}{dt}\equiv +g_{22}L_{12}^A \frac{m \Delta \dot{x}}{T}\;.
\end{eqnarray}
The first of the thermodynamic equations of motion as given in eqs. (\ref{Examples1.6}) yields
\begin{equation}
\label{Ex1.24}
-g_{11}\Delta x=+\frac{1}{(L_{12}^A)^2 g_{22}}\frac{d^2\Delta x}{dt^2}
\end{equation}
This is the thermodynamic analogy to Newton's second law, which is given by 
\begin{equation}\label{Ex1.25}
-K\Delta x = m\Delta\ddot{x}
\end{equation}
The thermodynamic formalism is absolutely equivalent to the analytical mechanical description of the same problem as given in eq. (\ref{T3.3}). One obtains oscillations of position and velocity of the piston, of internal energy and the flux of internal energy from one container to the other, and of the temperature of the two containers. 

\subsection*{LC circuit}

The derivations in the previous paragraph are independent of the choice of the extensive thermodynamic variable. 

Let us again consider two variables, $\xi_1$ and $\xi_2$. The first variable $\xi_1\equiv\Delta q$ shall be a charge, e.g. the difference of the charge on two capacitor plates. The second variable is given by
\begin{equation}
\label{LC1b}
\xi_2=-\frac{1}{L_{12}^A g_{22}}\frac{d \Delta q}{dt}=+\frac{1}{L_{12}^A g_{22}}I\;,
\end{equation}
where $I=-d\Delta q/dt$ is an electrical current.
\begin{figure}[hbt]
	\centering
	\includegraphics[width=6cm]{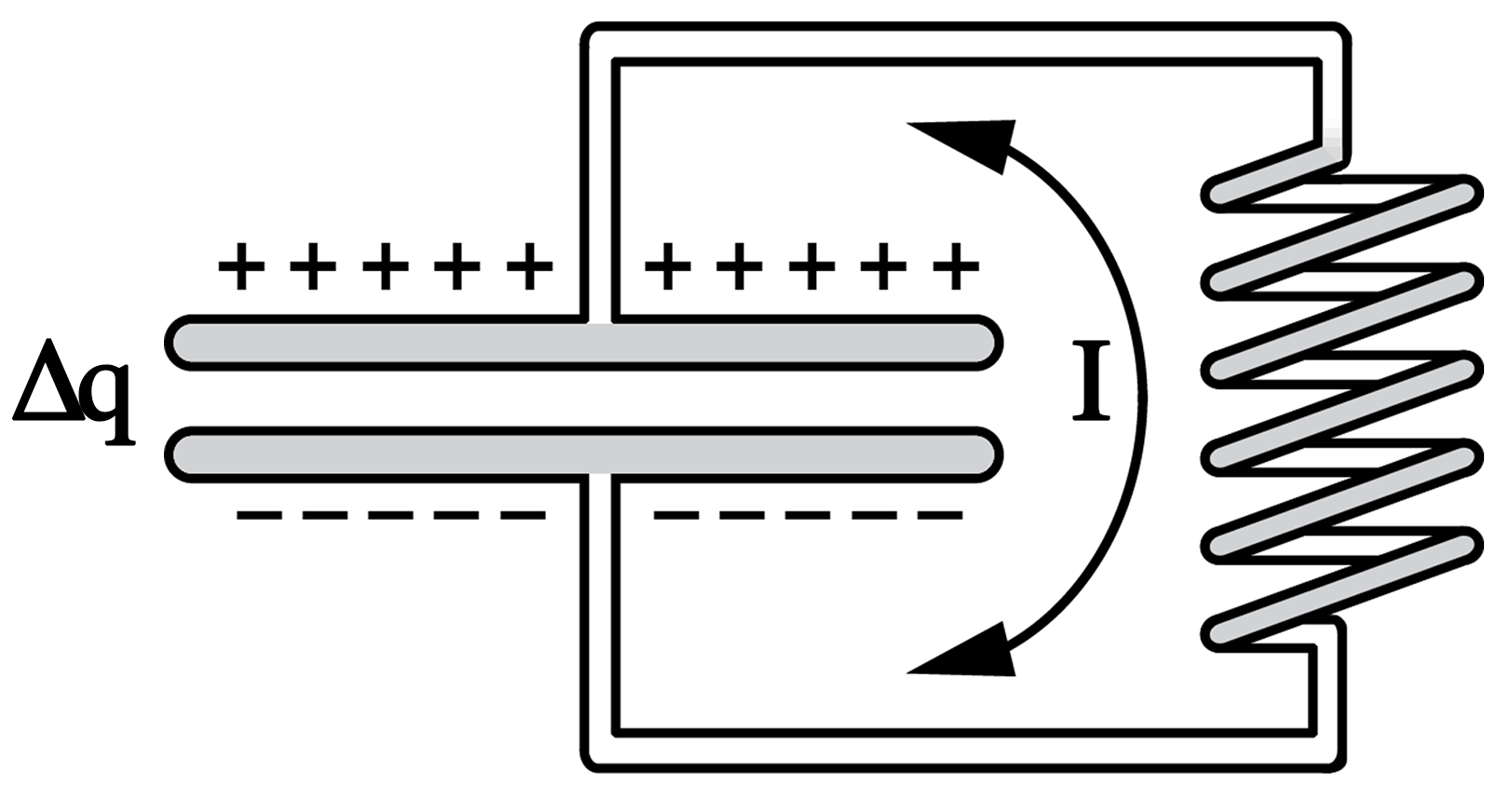}
	\caption{Schematic diagram of an LC circuit with a capacitance $C_m$ and the inductance of a solenoid coil of $L$. }
	\label{Figure_05}
\end{figure}
The entropy is conserved
\begin{equation}
\label{Figure2b3}
\Delta S=-\frac{1}{2}g_{11}\Delta q^2-\frac{1}{2g_{22} (L_{12}^A)^2}I^2=\mbox{const.}
\end{equation}
The analogous equation for an LC-circuit in electromagnetism is
\begin{equation}
\label{LC8}
E=\frac{1}{2}\frac {\Delta q^2}{C_m}+\frac{1}{2}LI^2=\mbox{const.} \;,
\end{equation}
where $C_m$ is the capacitance and $L$ is the inductance of the coil (Fig. \ref{Figure_05}). In analogy to the mechanical example, we find that $g_{11}\equiv 1/C_m T$ and $1/g_{22}(L_{12}^A)^2\equiv L/T$. The first term in eq. (\ref{LC8}) is the electric energy, while the second term is the magnetic energy. They correspond to the potential and the kinetic energy in the previous problem. In the language used here, eq. (\ref{LC8}) translates to
\begin{equation}
\label{LC8b}
-T\Delta S=\frac{1}{2}\frac {\Delta q^2}{C_m}+\frac{1}{2}LI^2=\mbox{const.} \;,
\end{equation}
The probability distribution for the charge fluctuations on the capacitor in equilibrium is then given by
\begin{equation}
\label{Ex1.19c2}
P_q=\sqrt{\frac{1}{2\pi kT C_m} } \cdot \exp\left( -\frac{\Delta q^2}{2 C_m kT}\right) 
\end{equation}
and the corresponding distribution for the current fluctuations (i.e., the Maxwell distribution) is
\begin{equation}
\label{Ex1.19d2}
P_{I}=\sqrt{\frac{L}{2\pi kT} }\cdot \exp\left( -\frac{L I^2}{2 kT}\right) 
\end{equation}
Eq. (\ref{Examples1.3}) yields
\begin{equation}
\label{LC8b2}
\frac{d^2 (\Delta q)}{dt^2} =  -\left(L_{12}^A\right)^2\left(g_{11}g_{22}\right)\Delta q
\end{equation}
with oscillatory solutions with frequency
\begin{equation}
\label{LC9}
\omega^2=\left(L_{12}^A\right)^2 g_{11} g_{22}=\frac{1}{LC_m} \;.
\end{equation}
The thermodynamic forces are given by
\begin{eqnarray}
\label{LC10}
X_1 & = &  -g_{11}\Delta x=-\frac{\Delta q}{C_m T}\equiv -\frac{\Psi_{el}}{T}\qquad\mbox{and} \nonumber \\
X_2 & = &  -g_{22}\xi_2=+\frac{1}{L_{12}^A}I\equiv +g_{22}L_{12}^A \frac{L\cdot I}{T}\;,
\end{eqnarray}
where $\Psi_{el}$ is the electrostatic potential of the capacitor. The first of the thermodynamic equations of motion as given in eq. (\ref{Examples1.6}) yields
\begin{equation}
\label{LC11}
-g_{11}\Delta q  =  +\frac{1}{(L_{12}^A)^2 g_{22}}\frac{d^2 \Delta q}{dt^2} \;.
\end{equation}
which corresponds to
\begin{equation}
\label{LC12}
-\frac{1}{C_m}\Delta q = L\dot{I}  \;.
\end{equation}
This equation has been derived previously from Kirchhoff's loop rule. Eq. (\ref{LC12}) is the electrical analogy to Newton's second law. $L\dot{I}=-L\Delta\ddot{q}$ is the equivalent of an inertial force, an $L$ plays the role of an inertial mass. Thus, the thermodynamic formalism is equivalent to the electrical description of the same problem.\\

For a solenoid with $N$ windings and a cross-section of $A$, eq. (\ref{LC12}) can also be written as
\begin{equation}
\label{LC13}
-\Psi_{el} = N\cdot A \frac{dB}{dt}
\end{equation}
Here, $\Psi_{el}=\Delta q/C_m$, $B=(\mu_0\mu_r N/l)\cdot I$, and the inductance is $L=\mu_0 \mu_r N^2 A/l$. The electrical potential $\Psi_{el}$ on the capacitor is equal to the electromotive force of the solenoid. This  well-known law  can also be derived from Faraday's law. 

It can be seen that in the above system one obtains oscillations of charge, electrostatic potential, current, magnetic field and temperature. The temperature oscillation will for instance be visible in the dielectric of the capacitor due to the electro-caloric effect \cite{Mischenko2006, Scott2011, Crossley2016} and in the diamagnetic or paramagnetic material inside of the coil (magneto-caloric effect). The charging of the capacitor and the magnetization of the coil without exchange of heat corresponds to the adiabatic compression of the ideal gas in the previous example.

\section*{Oscillating reactions}
The previous section suggests that there exist analogues of inertia also in systems that are not of mechanical nature. In the electromagnetic case, the inductance plays the role of an inertial mass. The results of the above calculations are well-known and therefore are not new. However, they demonstrate that these relations can be derived from a non-equilibrium formalism. The elementary constants in the formalism are already contained in the equilibrium fluctuations in the absence of any macroscopic motion. 

Since the different work terms all play analogous roles in thermodynamics, it seems plausible that any pair of an extensive variable and its conjugate intensive variable can be treated in a similar manner. This suggests that similarly meaningful thermodynamic relations can be obtained for any pair of variables. In the above examples, the mechanical work on a capacitor is $-\int F dx$, and the electrical work is $\int\Psi dq$. 

In chemistry, the intensive and extensive variables of interest are the chemical potentials, $\mu_i$, and the number of particles in a chemical reaction, $n_i$, and the work is given by $\int \sum \mu_i dn_i$. In a chemical reaction, the extensive quantity is the reaction variable, $\zeta$. 
Let us consider a chemical reaction
\begin{equation}
\label{CR0}
\nu_{X_1} X_1 + \nu_{X_2} X_2 + ... \stackrel{\zeta}{\leftrightarrow} \nu_{Y_1} Y_1 + \nu_{Y_2} Y_2 + ... \;,
\end{equation}
where the $X_i$ and $Y_i$ are chemical reagents, and the $\nu_i$ are the reaction stoichiometries. We assume that the first variable $\xi_{1}= \zeta$ is the reaction variable with the conjugated force $A/T$. Here, $A=-\left(\sum \nu_{Y_i} \mu_{Y_i}-\sum \nu_{X_i} \mu_{X_i}\right)$ is the affinity of the above reaction, the $\mu_i$ are the chemical potentials of the reagents, and the chemical work is given by $\int -A d\xi$. 

Most reactions are of a purely dissipative nature and do not display oscillations. However, there are many known chemical oscillations. Let us assume a chemical oscillation that is dominated by antisymmetric terms, i.e., dissipation is small. Then, the second variable $\xi_{2}$ is given by
\begin{equation}
\label{CR1}
\xi_{2}=-\frac{1}{L_{12}^A g_{22}}\frac{d \zeta}{dt} \;,
\end{equation}
where $(d\zeta/dt)$ is the flux in the chemical reaction.

The entropy is conserved
\begin{equation}
\label{CR2}
\Delta S=-\frac{1}{2}g_{11}\zeta^2-\frac{1}{2g_{22} (L_{12}^A)^2}\left(\frac{d\zeta}{dt}\right)^2=\mbox{const.} \;,
\end{equation}
where $1/g_{22}(L_{12}^A)^2\equiv L_{ch}/T $ and $g_{11}=1/C_{ch}T$. We call $L_{ch}$  a chemical inductance and $C_{ch}$ a chemical capacitance.

Eq. (\ref{Examples1.3}) yields
\begin{equation}
\label{CR3}
\frac{d^2 (\zeta)}{dt^2}  =  -\left(L_{12}^A\right)^2\left(g_{11}g_{22}\right) \zeta
\end{equation}
with oscillatory solutions with frequency
\begin{equation}
\label{CR4}
\omega^2=\left(L_{12}^A\right)^2 g_{11} g_{22}\equiv\frac{1}{C_{ch}L_{ch}} \;.
\end{equation}
The thermodynamic forces are given by
\begin{eqnarray}
\label{CR5}
X_1 & = &  -g_{11}\zeta=-\frac{\zeta}{C_{ch} T}\equiv \frac{A}{T}\qquad\mbox{and} \nonumber \\
X_2 & = &  -g_{22}\xi_2=+\frac{1}{L_{12}^A}\frac{d\zeta}{dt}\equiv +g_{22}L_{12}^A \frac{L_{ch}\cdot \dot{\zeta}}{T}\;,
\end{eqnarray}
with $-\zeta/C_{ch}=A$, where $A$ is the affinity.
The first of the thermodynamic equations of motion as given in eq. (\ref{Examples1.6}) yields
\begin{equation}
\label{CR6}
-g_{11}\zeta  =  +\frac{1}{(L_{12}^A)^2 g_{22}}\frac{d^2 \zeta}{dt^2} \;.
\end{equation}
which corresponds to
\begin{equation}
\label{CR7}
A = L_{ch}\ddot{\zeta}  \;,
\end{equation}
which is the chemical equivalent of Newton's second law.

As argued above, the chemical potentials, the number of particles of each chemical species, the affinity and the reaction variable $\zeta$ will oscillate. So will the temperature. The latter effect could be called a chemo-caloric effect. A shift in the position of a reaction without exchange of heat corresponds to the adiabatic compression of the ideal gas in the first example, and to the adiabatic charging of a capacitor in the second example. Temperature oscillations as indicator of the presence of adiabatic processes have been found in various chemical oscillations, most notably in the Belousov-Zhabotinsky reaction as discussed in the introduction (Fig. \ref{Figure_02}). We therefore propose that the oscillatory part of chemical clocks corresponds to the isentropic contribution to the chemical process.

\section{Discussion}
In equilibrium, ensembles do not display macroscopic motion. Therefore, any macroscopic oscillatory process necessarily describes a nonequilibrium situation. However, oscillations are not considered in the usual nonequilibrium thermodynamics formalism as given by Onsager \cite{Onsager1931a, Onsager1931b}, Prigogine and collaborators \cite{Kondepudi1998} or standard textbooks such as de Groot and Mazur \cite{deGroot1962}. We have shown here that by a generalization of the methods of linear non-equilibrium thermodynamics one can understand some simple oscillatory processes using the language of thermodynamics. While our considerations are very straight-forward and simple, they have (to our knowledge) not been made previously. Conceptually, however, the very insightful papers by von Helmholtz in his work on mono-cyclic systems \cite{Helmholtz1884} and on the principle of least action \cite{Helmholtz1887} already attempted to generalize the Lagrangian and Hamiltonian formalism from analytical mechanics to electrical systems and to thermodynamics. Helmholtz relates functions such as the Lagrangian to equivalent terms in thermodynamics such as the free energy. Here, we include the possibility that chemical oscillations can be understood by using such a formalism. 

A single oscillating mass as in Fig. \ref{Figure_01} is not an ensemble. The treatment of this problem with thermodynamic means is nevertheless possible because the springs are ensembles. Thus, thermodynamics provides the theory for the potential and the forces. It also provides a natural explanation why the mechanical spring possesses an equilibrium position. This is not required in an energy-based formalism. The origin of the potentials is not addressed in classical mechanics. This also holds true for the electrical example in Fig. \ref{Figure_05}.

Here, we focus on the linear regime of nonequilibrium thermodynamics where entropy production is given by $\sigma=\sum_i L_{ij}X_i X_j$. It is usually assumed that Onsager's reciprocal relations are valid, i.e., the coupling matrix is symmetric ($L_{ij}\equiv L_{ij}^S$) and the processes are purely dissipative. This formulation leads to interesting extremal laws such as minimum entropy production for stationary states \cite{Kondepudi1998}. It can be used to understand the couplings between different thermodynamic forces, e.g., Seebeck-, Peltier- , Dufour- and Soret-effect. Another principle is the maximum entropy production law \cite{Ziegler1961, Martyushev2006, Dewar2009}.  Maximum entropy production has been proposed to be important in the nonlinear physics of the atmosphere \cite{Dyke2010, Kleidon2010}, of evolution and life in general \cite{Martyushev2006, Michaelian2011}. However, the maximum entropy production principle is not well established and has been heavily criticized \cite{Ross2012}. It is thus unclear whether it represents a generic thermodynamic principle. 

There have been many attempts to include nonlinear processes to describe chemical oscillations \cite{Nicolis1971, Field1974}, complex systems \cite{Nicolis1977} and even the cyclic evolution of macromolecules and life \cite{Eigen1971}. Since life involves many chemical and biological clocks \cite{Rensing2001}, it is surprising that the focus in all of the above studies has been exclusively on the far-from-equilibrium dissipative processes. In contrast, the metaphor of a clock in physics describes mostly near-equilibrium reversible phenomena with only minor friction, which is compensated by the winding of a spring.

We show that the antisymmetric terms in Onsager's coupling matrix $L_{ij}$ do not contribute to entropy production. Martyushev and Seleznev \cite{Martyushev2006} concluded from this finding that Onsager's Matrix must be symmetric and antisymmetric contributions can be discarded. More radically, Coleman and Truesdell \cite{Coleman1960} questioned the general validity of the reciprocal relations. This is {\em not} the aim of the present paper. We agree that irreversible thermodynamics is described by a symmetric matrix. However, antisymmetric contributions provide additional information about co-existing reversible processes. 

Phenomenological Onsager-type equations yield fluxes that can be written as $\underline{J}=\underline{\underline{L}}\,\underline{X}$, where $\underline{\underline{L}}$ is an arbitrary matrix that can be uniquely separated into a symmetric and an antisymmetric matrix,  $\underline{\underline{L}}^S$ and  $\underline{\underline{L}}^A$, respectively. Any symmetric square matrix has only real eigenvalues and therefore leads to exponential relaxation behavior. It describes the irreversible processes. Any antisymmetric $n \times n$ matrix (where n is an even number) has only imaginary eigenvalues and leads to oscillatory solutions which do not contribute to dissipation.  The sum of a symmetric and an antisymmetric matrix can either have purely real eigenvalues or complex eigenvalues, depending on the magnitude of the antisymmetric matrix elements. 
Onsager's decision to focus on a symmetric coupling matrix \cite{Onsager1931b} is justified only in systems where motion is dominated by random thermal collisions, and inertia can be neglected. Under such conditions the correlation functions of thermal noise are independent of the arrow of time. This scenario was used for the derivation of the reciprocal relations. Even within these constraints the application of the reciprocal relations is not always trivial \cite{Casimir1945}. Onsager himself discussed  heat conduction in anisotropic crystals where heat flow in spirals may occur and the coupling matrix is not symmetric \cite{Onsager1931a}. Casimir discussed the example of heat conduction in asymmetric crystals, but also a case where a capacitor is discharged in the presence of magnetic fields where antisymmetric coefficients appear \cite{Casimir1945}. A similar case was discussed for static magnetic fields by Mazur and de Groot \cite{Mazur1953}.  Onsager's reciprocal relations require microscopic reversibility (detailed balance). Onsager writes in the abstract of his second 1931 paper \cite{Onsager1931b}: "(Quote) \textit{A general reciprocal relation, applicable to transport processes such as the conduction of heat and electricity, and diffusion, is derived from the assumption of microscopic reversibility.}" Thus, the argument that leads to the reciprocal relations is based on an assumption about the nature of microscopic subsystems in certain experimental settings. It is questionable whether this assumption is valid for macroscopic oscillations that include inertia, and where the arrow of time matters.

The example of two containers containing an ideal gas coupled by a piston with mass $m$ (Fig.\ref{Figure_01}) demonstrates that one can prepare thermodynamically meaningful nonequilibrium situations where Onsager's assumptions do not hold true. Oscillations represent nonequilibrium situations but they are not irreversible. We have shown here that one can find oscillations even close to equilibrium, and the far-from-equilibrium situations typically assumed for oscillating reactions are not necessary. For each extensive variable $\xi_1$ there is a variable $\xi_2=-(1/L_{12}^A g_{22})\dot{\xi_1}$ that behaves like a momentum. In the absence of oscillations, this momentum will decay with the same rate as the value of the extensive variable and no additional information is gained by considering the momentum. However, in the presence of inertia this process will include inertial forces that must be included in the formalism. With the present formalism, one can find analogies to the oscillation in a spring and the oscillations in an electrical LC element and find analogies to Newton's second law. In particular, one finds equivalents of the inertial mass. For instance, the mass associated with an oscillating spring is analogous to the inductance in a solenoid coil. Since the formalism is identical for each pair of extensive and intensive variables, we propose that isentropic oscillations in chemical reactions can exist in the linear thermodynamic limit, and that such oscillations are connected to chemical inertia involving the existence of a chemical inductance. In the linear limit, the thermodynamics forces are proportional to the extensive variables. Therefore, not only the extensive variables but also the intensive variables including the temperature fluctuate and oscillate. For this reason, the experimental finding of periodic temperature changes in chemical oscillations \cite{Franck1978, Franck1971, Koros1973, Boeckmann1996}) as well as the reversible temperature changes in nerve pulses \cite{Abbott1958, Howarth1968, Ritchie1985} are interesting and meaningful. It hints at adiabatic contributions to these oscillations. 

In a recent publication we argued that adiabatic processes are characterized by the reversible translation of energy between different parts of the system \cite{Mosgaard2013a}. This leads to measurable changes in temperature in the respective subsystems. Thus, each adiabatic oscillation requires a reservoir. In the case of the moving piston, the kinetic energy of the piston is stored in the energy of the ideal gas in the containers, which depends only on temperature. This leads to temperature oscillations. Therefore, including the energy translocation is essential for a complete understanding of the oscillatory processes. In the classical description of oscillatory reaction by coupled rate equations, this aspect is missing. As eq. (\ref{Intro0_d1}) shows, the entropy of an ideal gas contains two terms --- the first includes the changes in the motion of the particles and the second term is purely configurational entropy. The second leads to rate equations, because they are based on  statistical arguments about the positional distribution and the likelihood of collisions. Rates are considered to be proportional to concentrations only. The first part of eq. (\ref{Intro0_d1}) is not included in such rate equations and is therefore absent in reaction schemes leading to the Brusselator and the Oregonator \cite{Field1974, Nicolis1977, Kondepudi1998}. It would be interesting to investigate whether some terms in these reaction schemes can be reinterpreted such that they correspond to such energy reservoirs that reflect the temperature changes in the system.

The Brusselator and the Oregonator lead to stable limit cycles that are not very dependent on the initial conditions provided that some thermodynamic forces are kept constant. In contrast, in the present linear formalism forces are not fixed and therefore no limit cycles are present. However, depending on the relative contributions of the $L_{ij}^S$ and $L_{ij}^A$, the formalism can lead to purely dissipative behavior (real eigenvalues) and to bifurcations leading to dampened oscillations with a number of eigenfrequencies that depends on the dimension of the matrix $\underline{\underline{L}}$. In a future publication we will explore this in more detail. 

In statistical mechanics, pairs of a position and its corresponding momentum are considered for each particle. Liouville's theorem states that along the trajectories of a mechanical system the density of states in phase space is constant \cite{Sommerfeld1992e}. This is just another way of stating that entropy is conserved, which is precisely the assumption made here for oscillatory processes in a macroscopic system. Importantly, the constants of the oscillatory process are already hidden in the equilibrium fluctuations of position and momentum.

It is tempting to extend the simple examples presented here to more complex situations. Combining the mechanical and electrical examples given in this paper, one might find alternative formulations for phenomena related to piezoelectric oscillators. It is clear that the current $I$ couples to the mechanical motion in a piezoelectric crystal and that this leads to oscillations. It has been proposed that reversible electromechanical phenomena are at the basis of the action potential in nerves \cite{Heimburg2005c, GonzalezPerez2016}. Other phenomena include Rayleigh-B\'{e}nard convection. This phenomenon occurs when a liquid is homogeneously heated from below. Upon increasing the temperature gradients one finds bifurcations in the circular flow of the liquid and the formation of B\'{e}nard cells. In this system heat transfer couples to circular hydrodynamic motion. However, in the description of such phenomena the focus has been on entropy production on not so much on the entropy conserving parts \cite{Jia2014}, even though oscillatory and reversible solitary motion has been shown to be possible in Marangoni-B\'{e}nard layers \cite{Huang1998}. \\*{0.5cm}

\section*{Conclusions}
The postulate made in this work is that any spontaneous thermodynamic process in a harmonic entropy potential is composed of dissipative parts leading to entropy production and entropy conserving processes leading to oscillations. The thermodynamic formalism can be applied to any pair of variables and also to linear combinations of forces. We describe a simple formalism that can describe oscillatory processes by using the language of non-equilibrium thermodynamics. We suggest that any nonequilibrium process has a dissipative component described by Onsager's reciprocal relations and an oscillatory component that does not produce entropy. These processes are described by an antisymmetric coupling matrix. They are accompanied by oscillations in temperature which are characteristic for adiabatic waves. In fact, such temperature oscillations have been found in some chemical and biological systems. This suggests that they reflect isentropic processes contained in the overall reaction scheme. 
\section*{Acknowledgments}

This work was supported by the Villum Foundation (VKR 022130).  Prof. A. D. Jackson from the Niels Bohr International Academy, Prof. J. Ferkinghoff-Borg and Prof. B. Andresen from the University of Copenhagen and Prof. Lars Folke Olsen from the University of Southern Denmark critically read the manuscript and made valuable suggestions.

\small{

\begin{thebibliography}{10}
	
	\bibitem{Onsager1931a}
	Onsager, L.
	\newblock 1931.
	\newblock Reciprocal relations in irreversible processes. {I.}
	\newblock Phys.\ Rev. 37:405--426.
	
	\bibitem{Onsager1931b}
	Onsager, L.
	\newblock 1931.
	\newblock Reciprocal relations in irreversible processes. {II.}
	\newblock Phys.\ Rev. 38:2265--2279.
	
	\bibitem{Einstein1910}
	Einstein, A.
	\newblock 1910.
	\newblock Theorie der {O}paleszenz von homogenen {F}l{\"u}ssigkeiten und
	{F}l{\"u}ssigkeitsgemischen in der {N}{\"a}he des kritischen {Z}ustandes.
	\newblock Ann.\ Phys.\ (Leipzig) 33:1275--1298.
	
	\bibitem{Greene1951}
	Greene, R.~F., and H.~B. Callen.
	\newblock 1951.
	\newblock On the formalism of thermodynamic fluctuation theory.
	\newblock Phys.\ Rev. 83:1231--1235.
	
	\bibitem{Callen1952}
	Callen, H.~B., and R.~F. Greene.
	\newblock 1952.
	\newblock On a theorem of irreversible thermodynamics.
	\newblock Phys.\ Rev. 86:702--710.
	
	\bibitem{Heimburg1998}
	Heimburg, T.
	\newblock 1998.
	\newblock Mechanical aspects of membrane thermodynamics. {E}stimation of the
	mechanical properties of lipid membranes close to the chain melting
	transition from calorimetry.
	\newblock Biochim.\ Biophys.\ Acta 1415:147--162.
	
	\bibitem{Heimburg2012}
	Heimburg, T.
	\newblock 2012.
	\newblock The capacitance and electromechanical coupling of lipid membranes
	close to transitions. the effect of electrostriction.
	\newblock Biophys.\ J. 103:918--929.
	
	\bibitem{Zhabotinskii1964}
	Zhabotinskii, A.~M.
	\newblock 1964.
	\newblock Periodic oxidation reactions in liquid phase.
	\newblock Doklady Akademii Nauk SSSR 157:392--393.
	
	\bibitem{Bray1921}
	Bray, W.~C.
	\newblock 1921.
	\newblock A periodic reaction in homogeneous solution and its relation to
	catalyst.
	\newblock J.\ Am.\ Chem.\ Soc. 43:1262--1267.
	
	\bibitem{Briggs1973}
	Briggs, T.~S., and W.~C. Rauscher.
	\newblock 1973.
	\newblock Oscillating iodine clock.
	\newblock J.\ Chem.\ Education 50:496--496.
	
	\bibitem{Teusink1996}
	Teusink, B., C.~Larsson, J.~Diderich, P.~Richard, K.~{van Dam}, L.~Gustafsson,
	and H.~V. Westerhoff.
	\newblock 1996.
	\newblock Synchronized heat flux oscillations in yeast cell populations.
	\newblock J.\ Biol.\ Chem. 271:24442--24448.
	
	\bibitem{Franck1978}
	Franck, U.~F.
	\newblock 1978.
	\newblock Chemical oscillations.
	\newblock Angew.\ Chem.\ Int.\ Edit. 17:1--15.
	
	\bibitem{Edelson1979}
	Edelson, D., R.~M. Noyes, and R.~J. Field.
	\newblock 1979.
	\newblock Mechanistic details of the {B}elousov-{Z}habotinsky oscillations.
	{II.} {T}he organic-reaction subset.
	\newblock Int.\ J.\ Chem.\ Kin. 11:155--164.
	
	\bibitem{Vidal1980}
	Vidal, C., J.~C. Roux, and A.~Rossi.
	\newblock 1980.
	\newblock Quantitative measurement of intermediate species in sustained
	belousov-zhabotinsky oscillations.
	\newblock J.\ Am.\ Chem.\ Soc. 102:1241--1245.
	
	\bibitem{Nicolis1977}
	Nicolis, G., and I.~Prigogine, 1977.
	\newblock Self-organization in nonequilibrium systems: From dissipative
	structures to order through fluctuations.
	\newblock John Wiley {\&} {S}ons.
	
	\bibitem{Kondepudi1998}
	Kondepudi, D., and I.~Prigogine, 1998.
	\newblock Modern thermodynamics.
	\newblock Wiley, Chichester.
	
	\bibitem{Field1974}
	Field, R.~J., and R.~M. Noyes.
	\newblock 1974.
	\newblock Oscillations in chemical systems. {IV}. {L}imit cycle behavior in a
	model of a real chemical reaction.
	\newblock J.\ Am.\ Chem.\ Soc. 60:1877--1884.
	
	\bibitem{Boeckmann1996}
	B\"ockmann, M., B.~Hess, and S.~M\"uller.
	\newblock 1996.
	\newblock Temperature gradients traveling with chemical waves.
	\newblock Phys.\ Rev.\ E 53:5498--5501.
	
	\bibitem{Franck1971}
	Franck, U., and W.~Geiseler.
	\newblock 1971.
	\newblock Zur periodischen {R}eaktion von {M}alons{\"a}ure mit {K}aliumbromat
	in {G}egenwart von {C}er-{I}onen.
	\newblock Naturwissenschaften 58:52--53.
	
	\bibitem{Koros1973}
	Orban, E. K.~M., and Z.~Nagy.
	\newblock 1973.
	\newblock Periodic heat evolution during temporal chemical oscillations.
	\newblock Nature 242:30--31.
	
	\bibitem{Lamprecht1987}
	Lamprecht, I., B.~Schaarschmidt, and T.~Plesser.
	\newblock 1987.
	\newblock Heat-production in oscillating chemical-reactions - 3 examples.
	\newblock Thermochim.\ Acta 112:95--100.
	
	\bibitem{Abbott1958}
	Abbott, B.~C., A.~V. Hill, and J.~V. Howarth.
	\newblock 1958.
	\newblock The positive and negative heat production associated with a nerve
	impulse.
	\newblock Proc.\ Roy.\ Soc.\ Lond.\ B 148:149--187.
	
	\bibitem{Ritchie1985}
	Ritchie, J.~M., and R.~D. Keynes.
	\newblock 1985.
	\newblock The production and absorption of heat associated with electrical
	activity in nerve and electric organ.
	\newblock Quart.\ Rev.\ Biophys. 18:451--476.
	
	\bibitem{Heimburg2005c}
	Heimburg, T., and A.~D. Jackson.
	\newblock 2005.
	\newblock On soliton propagation in biomembranes and nerves.
	\newblock Proc.\ Natl.\ Acad.\ Sci.\ USA 102:9790--9795.
	
	\bibitem{GonzalezPerez2016}
	{Gonzalez-Perez}, A., L.~D. Mosgaard, R.~Budvytyte, E.~{Villagran Vargas},
	A.~D. Jackson, and T.~Heimburg.
	\newblock 2016.
	\newblock Solitary electromechanical pulses in lobster neurons.
	\newblock Biophys.\ Chem. 216:51--59.
	
	\bibitem{Kubo1966}
	Kubo, R.
	\newblock 1966.
	\newblock The fluctuation-dissipation theorem.
	\newblock Rep.\ Prog.\ Phys. 29:255--284.
	
	\bibitem{Coleman1960}
	Coleman, B.~D., and C.~Truesdell.
	\newblock 1960.
	\newblock On the reciprocal relations of onsager.
	\newblock J.\ Chem.\ Phys. 33:28--31.
	
	\bibitem{Martyushev2006}
	Martyushev, L.~M., and V.~D. Seleznev.
	\newblock 2006.
	\newblock Maximum entropy production principle in physics, chemistry and
	biology.
	\newblock Physics Reports 426:1--45.
	
	\bibitem{Mischenko2006}
	Mischenko, A.~S., Q.~Zhang, J.~F. Scott, R.~W. Whatmore, and N.~D. Mathur.
	\newblock 2006.
	\newblock Giant electrocaloric effect in thin-film
	{P}b{Z}r$_{0.95}${T}i$_{0.05}${O}$_3$.
	\newblock Science 311:1270--1271.
	
	\bibitem{Scott2011}
	Scott, J.~F.
	\newblock 2011.
	\newblock Electrocaloric materials.
	\newblock Annu.\ Rev.\ Mater.\ Res. 41:229--240.
	
	\bibitem{Crossley2016}
	Crossley, S., T.~Usui, B.~Nair, S.~Kar-Narayan, X.~Moya, S.~Hirose, A.~Ando,
	and N.~D. Mathur.
	\newblock 2016.
	\newblock Direct electrocaloric measurement of
	0.9{P}b({M}g$_{1/3}${N}b$_{2/3}$){O}$_3$-0.1{P}b{T}i{O}$_3$ films using
	scanning thermal microscopy.
	\newblock Appl.\ Phys.\ Letters 108:032902.
	
	\bibitem{deGroot1962}
	{de Groot}, S.~R., and P.~Mazur, 1962.
	\newblock Non-equilibrium thermodynamics.
	\newblock North-{H}olland {P}ublishing Co., Amsterdam.
	
	\bibitem{Helmholtz1884}
	{von Helmholtz}, H.
	\newblock 1884.
	\newblock Principien der statik monocyklischer systeme.
	\newblock J.\ Reine\ Angew.\ Math. 97:111--140.
	
	\bibitem{Helmholtz1887}
	{von Helmholtz}, H.
	\newblock 1887.
	\newblock \"{U}ber die physikalische {B}edeutung des {P}rincips der kleinsten
	{W}irkung.
	\newblock Journal f\"{u}r die {R}eine und {A}ngewandte {M}athematik
	100:137--166.
	
	\bibitem{Ziegler1961}
	Ziegler, H.
	\newblock 1961.
	\newblock Zwei {E}xtremalprinzipien der irreversiblen {T}hermodynamik.
	\newblock Ingenieur-{A}rchiv 30:410--416.
	
	\bibitem{Dewar2009}
	Dewar, R.~C.
	\newblock 2009.
	\newblock Maximum entropy production as an inference algorithm that translates
	physical assumptions into macroscopic predictions: {D}on't shoot the
	messenger.
	\newblock Entropy 11:931--944.
	
	\bibitem{Dyke2010}
	Dyke, J., and A.~Kleidon.
	\newblock 2010.
	\newblock The maximum entropy production principle: Its theoretical foundations
	and applications to the earth system.
	\newblock Entropy 12:613--630.
	
	\bibitem{Kleidon2010}
	Kleidon, A.
	\newblock 2010.
	\newblock Life, hierarchy, and the thermodynamic machinery of planet earth.
	\newblock Phys.\ Life Rev. 7:424--460.
	
	\bibitem{Michaelian2011}
	Michaelian, K.
	\newblock 2011.
	\newblock Entropy production and the origin of life.
	\newblock J.\ Mod.\ Phys. 2:595--601.
	
	\bibitem{Ross2012}
	Ross, J., A.~D. Corlan, and S.~C. M\"uller.
	\newblock 2012.
	\newblock Proposed principles of maximum local entropy production.
	\newblock J.\ Phys.\ Chem.\ B 116:7858--7865.
	
	\bibitem{Nicolis1971}
	Nicolis, G.
	\newblock 1971.
	\newblock Stability and dissipative structures in open systems far from
	equilibrium.
	\newblock Adv.\ Chem.\ Phys. 19:209--324.
	
	\bibitem{Eigen1971}
	Eigen, M.
	\newblock 1971.
	\newblock Selforganization of matter and evolution of biological
	macromolecules.
	\newblock Naturwissenschaften 58:465--523.
	
	\bibitem{Rensing2001}
	Rensing, L., U.~{Meyer-Grahle}, and P.~Ruoff.
	\newblock 2001.
	\newblock Biological timing and the clock metaphor: {O}scillatory and hourglass
	mechanicms.
	\newblock Chronobiology Int. 18:329--369.
	
	\bibitem{Casimir1945}
	Casimir, H. B.~G.
	\newblock 1945.
	\newblock On {O}nsager's principle of microscopic reversibility.
	\newblock Rev.\ Mod.\ Phys. 17:343--350.
	
	\bibitem{Mazur1953}
	Mazur, P., and S.~R. {de Groot}.
	\newblock 1953.
	\newblock On {O}nsager's relations in a magnetic field.
	\newblock Physica 19:961--970.
	
	\bibitem{Howarth1968}
	Howarth, J.~V., R.~Keynes, and J.~M. Ritchie.
	\newblock 1968.
	\newblock The origin of the initial heat associated with a single impulse in
	mammalian non-myelinated nerve fibres.
	\newblock J.\ Physiol. 194:745--793.
	
	\bibitem{Mosgaard2013a}
	Mosgaard, L.~D., A.~D. Jackson, and T.~Heimburg.
	\newblock 2013.
	\newblock Fluctuations of systems in finite heat reservoirs with applications
	to phase transitions in lipid membranes.
	\newblock J.\ Chem.\ Phys. 139:125101.
	
	\bibitem{Sommerfeld1992e}
	Sommerfeld, A., 1992.
	\newblock Thermodynamik und Statistik, volume~5 of \emph{Vorlesungen \"uber
		theoretische {Physik}}.
	\newblock Harri Deutsch.
	
	\bibitem{Jia2014}
	Jia, C., C.~C.~Jing, and J.~Liu.
	\newblock 2014.
	\newblock The character of entropy production in {R}ayleigh--{B}{\'e}nard
	convection.
	\newblock Entropy 16:4960--4973.
	
	\bibitem{Huang1998}
	Huang, G.~X., M.~G. Velarde, and V.~N. Kurdyumov.
	\newblock 1998.
	\newblock Cylindrical solitary waves and their interaction in benard-marangoni
	layers.
	\newblock Phys.\ Rev.\ E 57:5473--5482.
	
\end{thebibliography}

}

\end{document}